\documentclass[aps,prc,twocolumn,groupedaddress,showpacs]{revtex4}

\usepackage{graphicx}

\begin{document}

\title{Symmetry energy and the isospin dependent equation of state}
\author{D.V. Shetty, S.J. Yennello, A.S. Botvina\footnote{On leave from Institute for Nuclear Research, 117312 Moscow, Russia.}, 
          G.A. Souliotis, M. Jandel, E. Bell, A. Keksis, S. Soisson, B. Stein, and J. Iglio}
\affiliation{Cyclotron Institute, Texas A$\&$M University, College Station, Texas 77843, USA}
\date{\today}

\begin{abstract}
The isoscaling parameter $\alpha$, from the fragments produced in the multifragmentation of $^{58}$Ni + $^{58}$Ni, 
$^{58}$Fe + $^{58}$Ni and $^{58}$Fe + $^{58}$Fe reactions at 30, 40 and 47 MeV/nucleon, was 
compared with that predicted by the antisymmetrized molecular dynamic (AMD) calculation based on two 
different nucleon-nucleon effective forces, namely the Gogny and Gogny-AS interaction. The results show that the 
data agrees better with the choice of Gogny-AS effective interaction, resulting in a symmetry energy of 
$\sim$18 - 20 MeV.  The observed value indicates that the fragments are formed at a reduced density of $\sim$ 0.08 fm$^{-3}$.
\end{abstract}


\maketitle

\section{Introduction}
The asymmetry term in the nuclear equation of state (EOS) of strongly interacting matter is important for studying the structure, 
chemical compositions and evolution of neutron stars, and the dynamics of supernova explosion \cite{LAT91, LEE96, BET90}. 
This term dominates the pressure within the neutron stars at densities of $\rho$ $\leq$ 2$\rho_{\circ}$ (where $\rho_{\circ}$ is 
the normal matter nuclear density), and modifies the protoneutron star cooling rates \cite{PET95, LAT00}. In a supernova 
explosion, the variation of the asymmetry energy can also change the rate of electron capture in a collapsing star, thereby 
altering the course of final explosion \cite{HIX03}. Under laboratory controlled conditions, the low density dependence of the 
asymmetry term in EOS can be investigated in a multifragmentation reaction, where the system expands to a density of  $\rho$  
$\approx$ $\rho_{\circ}/6$ - $\rho_{\circ}/3$ and disassembles into various light and heavy fragments \cite{BOR01, BOW91, AGO96, BEA00}. 
The isotopic composition of these fragments determines the properties of the system and has significant sensitivity to the 
density dependence of the asymmetry term. Various theoretical studies \cite{BAO97, BAO00, TAN01, TSA01} have explored 
this influence of the density dependence of the asymmetry term,  however, the detailed nature of this dependence is model dependent. 
For example, the hybrid model approach predicts that an ``asy-stiff" EOS leads to fragments that are more neutron-rich than 
those produced when the EOS is  ``asy-soft" \cite{TAN01}. On the other hand, calculations with the expanding evaporating 
source (EES) model predict the opposite trend \cite{TSA01}. 
\par
Recently \cite{ONO03},  the fragment yields from the nuclear collisions simulated within the antisymmetrized molecular dynamics 
(AMD) were reported to follow a scaling behavior of the type,
\begin{equation}
     Y_{2}(N,Z)/Y_{1}(N,Z) \propto e^{\alpha N + \beta Z}
\end{equation}
where the parameters $\alpha$ and $\beta$ are related to the neutron-proton content of the fragmenting source. 
A linear relation between the isoscaling parameter $\alpha$, and the isospin 
asymmetry (Z$/$A)$^{2}$ of the fragments, with appreciably different slopes, was observed for two different choices of the effective 
nucleon-nucleon interaction force.  It was shown that such a relation can be used to limit the symmetry energy parameter in the nuclear 
equation of state and extract density pertinent to the fragmentation.  
\par
In this paper, we compare the experimentally observed scaling parameter $\alpha$, to those predicted by two different 
effective nucleon-nucleon interaction forces, namely the Gogny and Gogny-AS (a modified form of the Gogny interaction), in the 
AMD calculation.  It is shown that the Gogny-AS interaction explains the observed results better. The symmetry energy is found to be 
about 18 - 20 MeV, indicating that the fragments are formed when the system expand to a density $\rho$  $\sim$ 0.08 fm$^{-3}$.
\section{Experiment}
The measurements were performed at the Cyclotron Institute of Texas A$\&$M University (TAMU) using beams provided by the K500
Superconducting Cyclotron. Isotopically pure beams of $^{58}$Ni and $^{58}$Fe at 30, 40 and 47 MeV/nucleon were bombarded on self-supporting 
$^{58}$Ni (1.75 mg/cm$^{2}$) and $^{58}$Fe (2.3 mg/cm$^{2}$) targets. The targets were placed in the center of a scattering chamber 
that was housed inside the TAMU 4$\pi$ neutron ball detector \cite{SCH95}. Six discrete particle telescopes placed inside a scattering chamber 
and centered at laboratory angles of 10$^{\circ}$, 44$^{\circ}$, 72$^{\circ}$, 100$^{\circ}$, 128$^{\circ}$ and 148$^{\circ}$, were used 
to measure fragments from the reactions. Each telescope consisted of a gas ionization chamber (IC) followed by a pair of silicon detectors 
(Si-Si) and a CsI scintillator detector, providing three distinct detector pairs (IC-Si, Si-Si, and Si-CsI) for fragment identification. The 
    \begin{figure}
    \includegraphics[width=0.45\textwidth,height=0.45\textheight]{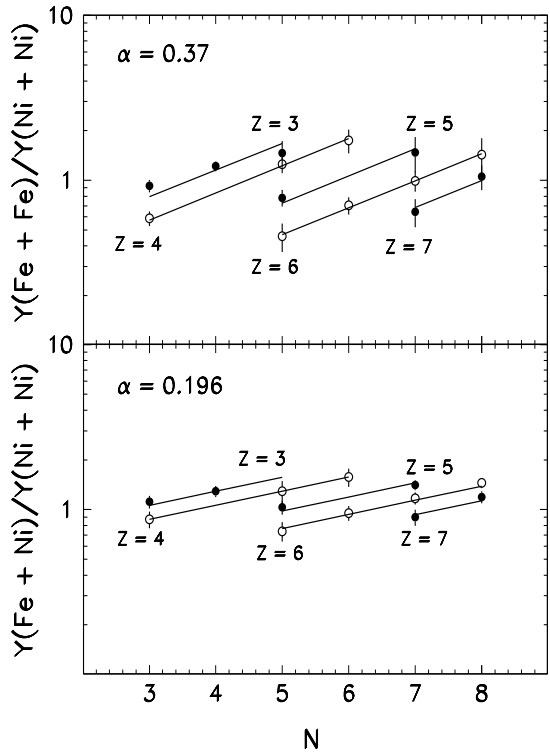} 
    \caption{Yield ratios of the fragments from Fe + Fe and Ni + Ni reactions (top), and Fe + Ni 
                 and Ni + Ni reactions (bottom), as a function of neutron number N for the 30 MeV/nucleon beam energy. 
                 The solid lines are the exponential fits to the data. The fit parameter $\alpha$, is shown in the top left corner of each panel.}
   \end{figure}
ionization chamber was of axial field design and was operated with CF$_{4}$ gas at a pressure of 50 Torr. The gaseous medium was 6 cm
thick with a typical threshold of $\sim$ 0.5 MeV/nucleon for intermediate mass fragments. The silicon detectors had an active area of 5 cm $\times$
5 cm and were each subdivided into four quadrants. The first and second silicon detectors in the stack were 0.14 mm and 1mm thick, respectively.
The dynamic energy range of the silicon pair was  $\sim$ 16 - 50 MeV for $^{4}$He and  $\sim$ 90 - 270 MeV for $^{12}$C. The CsI
scintillator crystals that followed the silicon detector pair were 2.54 cm in thickness and were read out by photodiodes.  Good Z identification 
was achieved for fragments that punched through the IC detector and stopped in the first silicon detector. Fragments measured in the Si-Si 
detector pair also had good isotopic separation. In this work, we present data collected in the detector telescope placed at 44$^{\circ}$ in the 
laboratory. The fragments detected at this angle originate predominantly from central events. The centrality was assured by gating on the 
measured neutron multiplicity. Further details can be found in Ref. \cite{RAM98}
\section{Results and Discussion}
Fig. 1 shows the yield ratios for the fragments with 3 $\leq$ Z $\leq$ 7 from $^{58}$Fe + $^{58}$Ni, 
$^{58}$Fe + $^{58}$Fe and $^{58}$Ni + $^{58}$Ni reactions at 30 MeV/nucleon, as a function of  neutron number N. The 
upper panel shows the ratios for the $^{58}$Fe + $^{58}$Fe and $^{58}$Ni + $^{58}$Ni reactions,  and the lower one for the
$^{58}$Fe + $^{58}$Ni and $^{58}$Ni + $^{58}$Ni reactions. The measured ratios clearly satisfy the scaling relation of the form 
shown in Eq. 1, and are shown by the solid lines in the figure. The scaling parameter $\alpha$ is as shown in each panel of Fig. 1.  
    \begin{figure}
    \includegraphics[width=0.5\textwidth,height=0.40\textheight]{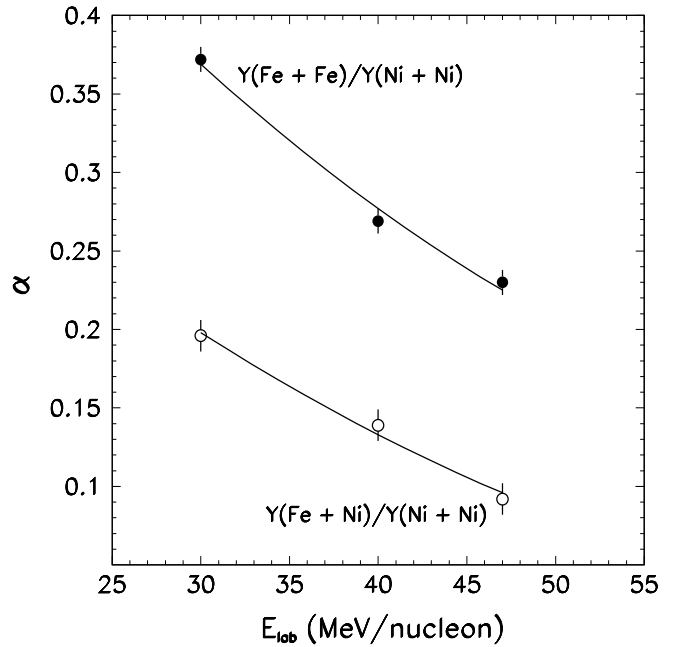} 
    \caption{Scaling parameter $\alpha$, obtained from a fit similar to those shown in Fig. 1, as a function of
                 bombarding energy. The solid lines are exponential fit to the data.}
   \end{figure}
A similar relation is also observed for the 40 and 47 MeV/nucleons beam energies. The values of  the $\alpha$ parameters 
from all three beam energies for the two different combinations of reactions are as shown in Fig. 2. The solid lines
are the exponential fits to the data. One sees a gradual decrease in the $\alpha$ values as expected due to an increase in the 
excitation energy or the temperature \cite{SHE03}.
\par
As previously discussed and shown in Ref. \cite{ONO03, BOT02}, the parameter $\alpha$, is related to the fragment isospin 
asymmetry, (Z$/$A)$^{2}$, through a linear relation of the form

\begin{equation}
      \alpha = \frac{4C_{sym}}{T} {[(Z/A)_{1}^{2} - (Z/A)_{2}^{2}]}
\end{equation}

where C$_{sym}$ is the symmetry energy and T is the temperature at which the fragments are formed. The quantities,  $(Z/A)_{i}^{2}$, 
are the isospin asymmetries of the fragments in the two reaction systems, ${\it {i}}$ = 1 and ${\it {i}}$ = 2. 
Based on the calculations predicted in Ref. \cite{ONO03}, the isospin asymmetry of the fragments for the present systems were 
estimated at t = 300 fm/c of the dynamical evolution. The asymmetry values, (Z/A)$^{2}$, at t = 300 fm/c for the present systems were 
obtained by interpolating between those of Ref. \cite{ONO03}.
    \begin{figure}
    \includegraphics[width=0.5\textwidth,height=0.40\textheight]{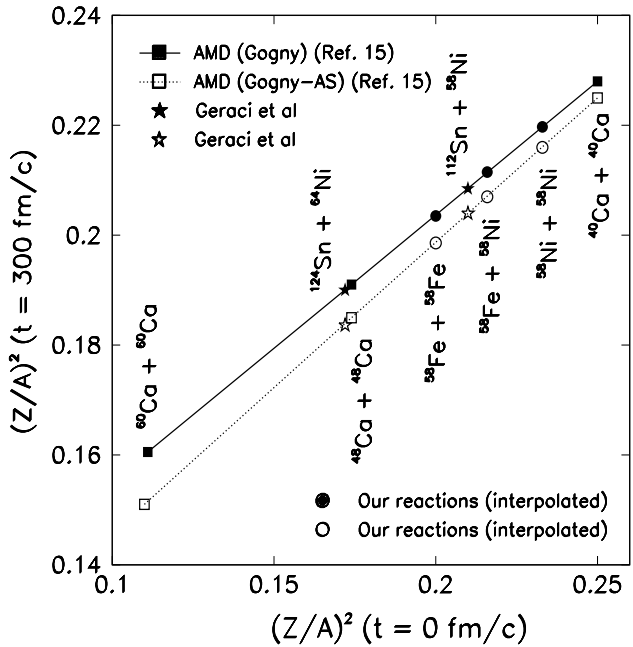} 
    \caption{AMD calculations of the asymmetry (Z/A)$^{2}$, of the fragments at t = 300 fm/c for the Gogny (solid line and solid squares) 
                 and Gogny-AS (dotted line and hollow squares) interactions at 35 MeV/nucleon. The calculations are taken from Ref. \cite{ONO03} 
                 for the systems shown by the square symbols. The lines are linear fit to the square symbols. The circle symbols are the 
                 interpolated values for the systems studied in this work.}
   \end{figure}
Fig. 3 shows the AMD calculation of the fragment asymmetry,   (Z/A)$^{2}$ at t = 300 fm/c, as a function of initial asymmetry at time 
t = 0 fm/c, for  two different choices of the nucleon-nucleon interaction force, Gogny and Gogny-AS. The asymmetry values for the
systems investigated in Ref. \cite{ONO03} are shown by the solid and hollow square symbols for the Gogny and Gogny-AS interaction, 
respectively. The lines are the linear fits to the calculations. The interpolated values for the present systems are shown by the solid 
and hollow circles. The solid and the hollow stars correspond to the reactions studied by Geraci {\it {et al.}} \cite{GER04}, and will be discussed
later. 
\par
It should be mentioned that the AMD calculations carried out in Ref. \cite{ONO03} and shown in Fig. 3 correspond to the beam energy of 35 MeV/nucleon.
The interpolated values of the asymmetries for the present systems obtained from Fig. 3 are thus for the beam energy of 35 MeV/nucleon.
In order to make a straightforward comparison, we obtain the $\alpha$ parameter for the present system at 35 MeV/nucleon by interpolating 
the observed values using Fig. 2. We then explore the relation shown in Eq. 2 and determine the symmetry energy, knowing the temperature T.  
\par
The temperatures for the present system were determined from the double isotope ratio method of Albergo {\it {et al.}} \cite{ALB85} for the 30 and 
40 MeV/nucleon and were observed to be 3.4 and 3.7 MeV respectively. This is consistent with the value of 3.4 MeV quoted in 
Ref. \cite{ONO03} for the systems of similar mass and energy.
\par
Fig. 4 shows the $\alpha$ parameter plotted as a function of the difference in the asymmetry of the fragments for the two reactions 
at t = 300 fm/c. The solid and the dotted lines are the AMD predictions of the Gogny and Gogny-AS effective forces at a beam 
energy of 35 MeV/nucleon, respectively.
    \begin{figure}
    \includegraphics[width=0.5\textwidth,height=0.40\textheight]{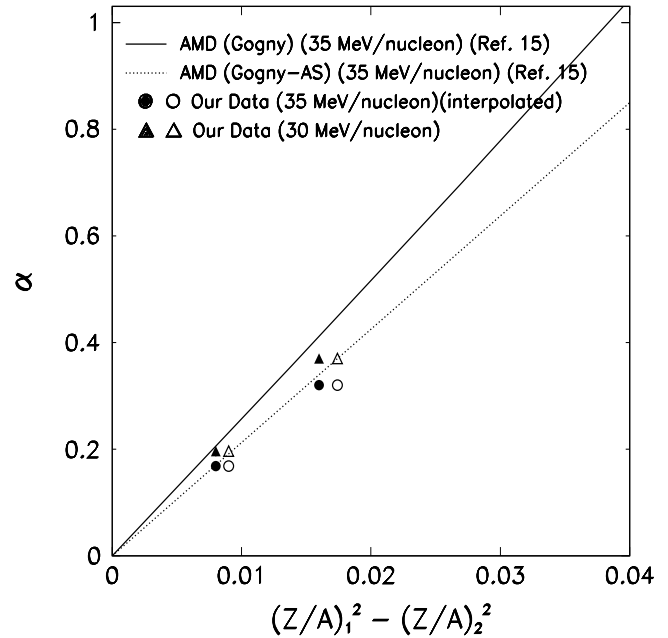} 
    \caption{Scaling parameter $\alpha$, as a function of the asymmetry of the fragments. The solid and the dotted lines are the AMD 
      calculations for the Gogny and Gogny-AS interactions at 35 MeV/nucleon, respectively \cite{ONO03}. The circle symbols are the 
      interpolated values of the parameter from the present work for 35 MeV/nucleon. The triangles are the observed values for the 
      30 MeV/nucleon. The error bars are of the size of symbols.}
   \end{figure}
    \begin{figure}
    \includegraphics[width=0.5\textwidth,height=0.40\textheight]{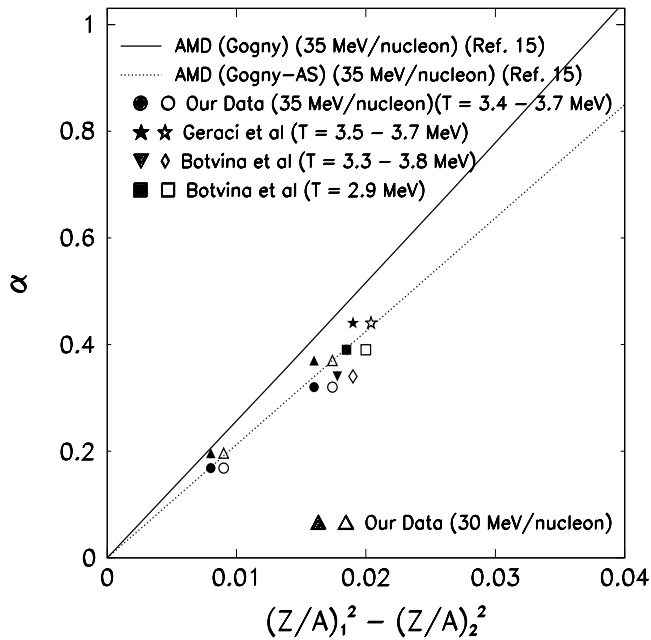} 
    \caption{Scaling parameter $\alpha$, as a function of the asymmetry of the fragments. The solid and the dotted lines are the AMD 
      calculations for the Gogny and Gogny-AS interactions at 35 MeV/nucleon, respectively \cite{ONO03}. The solid and hollow circle 
      symbols are the extracted values from the present work for 35 MeV/nucleon. The various symbols are as explained in the text.}
   \end{figure}
The solid and the hollow circles are the interpolated values obtained from the present measurements for the 35 MeV/nucleon using 
two different values of the difference in asymmetry, resulting  from the Gogny and Gogny-AS effective forces. For reference, the measured 
values of the $\alpha$ parameter for the 30 MeV/nucleon values are also shown in the figure by the solid and hollow triangles. As 
shown in the figure, the magnitude of the $\alpha$ parameter for the 35 MeV/nucleon linearly increases with increasing difference in 
the asymmetry of the two systems as predicted by the AMD calculation. Also, the parameter values are in better agreement with the 
values obtained from the Gogny-AS interaction than those from the usual Gogny force. A slightly lower value of the circle symbols for 
35 MeV/nucleon compared to the Gogny-AS value (dotted line) could be due to the secondary de-excitation effect, which can 
significantly lower the experimentally observed value \cite{SHE04}. The closer agreement with the Gogny-AS type of interaction 
nevertheless indicates a stiffer EOS relative to the usual Gogny interaction in the low density region of the EOS. A similar conclusion 
was also arrived at in a model dependent hybrid calculations for the $^{112}$Sn + $^{112}$Sn, $^{112}$Sn + $^{124}$Sn and 
$^{124}$Sn + $^{124}$Sn reactions at 50 MeV/A \cite{TAN01}.    
\par
From the slope, 4C$_{sym}$/T (= 21.2), of the Gogny-AS interaction (dotted line) in Fig. 4, and temperature T ( = 3.4 - 3.7 MeV), we 
obtain a value for symmetry energy C$_{sym}$ = 18 - 20 MeV. This suggests (see Fig. 1 of Ref. \cite{ONO03}) that fragments are 
formed at a density of $\rho$ $\sim$ 0.08 fm$^{-3}$, consistent with the fact that fragmentation occurs at a reduced density. A similar 
value of the symmetry energy C$_{sym}$ = 17 - 21 MeV, for the present system was also obtained following the 
statistical model calculation of Botvina {\it {et al.}} \cite{BOT02} which includes the secondary de-excitation of the primary fragments.
\par
We next compare the scaling parameter $\alpha$, taken from various works \cite{GER04, BOT02} with those obtained from the 
present study. The data were specifically chosen from reactions where the system temperature ranges from T = 3 MeV to T = 3.8 MeV, 
similar to those observed in the present study and obtained by the double isotope ratio method. The comparison is shown in Fig. 5, where 
the circle and the triangle symbols are the same as shown in Fig. 4 and correspond to the present study.
The solid and the hollow stars correspond to $^{124, 112}$Sn + $^{64, 58}$Ni reactions at 35 MeV/nucleon studied by 
Geraci {\it {et al.}} \cite{GER04}. The solid and the hollow squares are for the H + $^{124, 112}$Sn reactions at 6.7 GeV, whereas the 
inverted triangle and the diamond symbols are for the $^{4}$He + $^{124, 112}$Sn reactions at 15.3 GeV \cite{BOT02}.  All the data 
points shown correspond to reactions with nearly the same temperature as indicated in Fig. 5. One observes that the data points all lie 
scattered close to the prediction of Gogny-AS interaction (dotted line). The present observation thus indicates that at the threshold of 
multifragmentation (T = 3 - 3.8 MeV), the fragments are formed at a reduced density of  $\sim$ 0.08 fm$^{-3}$ with a symmetry energy 
C$_{sym}$ = 18 - 20 MeV, and the process is consistent with an equation of state determined by the Gogny-AS type of interaction force. 
A more detailed study using neutron rich beams and populating higher asymmetry region in the $\alpha$ versus fragment asymmetry 
plot, where the difference between the predictions for the Gogny and Gogny-AS interaction becomes large, could be important.
\section{Conclusion}
In conclusion, we have shown using experimental data that the linear dependence of the isoscaling parameter $\alpha$, as a 
function of fragment isospin asymmetry (Z/A)$^{2}$, is sensitive to  the symmetry terms used in the nuclear equation of state 
and follows a Gogny-AS interaction in the low density region of the EOS. Furthermore, at a temperature of T = 3 - 3.8 MeV, the fragments 
are formed at a reduced density of $\sim$ 0.08 fm$^{-3}$ with a symmetry energy of the order of 18 - 20 MeV.
\section{Acknowledgment}
The authors wish to thank the staff of the Texas A$\&$M Cyclotron facility for the excellent beam quality. This work was supported in 
part by the Robert A. Welch Foundation through grant No. A-1266, and the Department of Energy through grant No. DE-FG03-93ER40773. 
One of the authors (ASB) thanks Cyclotron Institute TAMU for hospitality and support. 

\bibliography{symmetry_paper.bbl}

\end{document}